\documentstyle[amsfonts,amssymb,amsbsy,aps,eqsecnum,psfig,floats,graphicx]{revtex}

\newcommand\gbar{\overline{g}}
\renewcommand\hbar{\overline{h}}
\newcommand\half{{\textstyle\frac{1}{2}}}
\newcommand\pstar{p^\star}
\newcommand\ansatz{{ansatz}}
\newcommand\gammahat{j}

\begin{document}

\title{Black-hole threshold solutions in stiff fluid collapse}
\author{Patrick R.\ Brady}
\address{Institute for Theoretical Physics, University of California
at Santa Barbara, Santa Barbara, CA 93106}
\address{Department of Physics, University of Wisconsin at Milwaukee,
PO Box 413, Milwaukee, WI 53201 \footnote{current address}}
\author{Matthew W.\ Choptuik}
\address{CIAR Cosmology and Gravity Program \\
Department of Physics and Astronomy,
University of British Columbia, 6224 Agricultural Road, 
Vancouver, V6T 1Z1, Canada}
\author{Carsten Gundlach}
\address{Enrico Fermi Institute, University of
Chicago, 5640 Ellis Avenue, Chicago, IL 60637}
\address{Faculty of Mathematical Studies, University of Southampton,
Southampton SO17 1BJ, UK \footnote{current address}}
\author{David W.\ Neilsen}
\address{Center for Relativity, The University of Texas at Austin,
Austin, TX 78712-1081}
\date{24 July 2002}
\maketitle

%%%%%%%%%%%%%%%%%%%%%%%%%%%%%%%%%%%%%%%%%%%%%%%%%%%%%%%%%%%%%%%%%%%%%%%%%%%
\begin{abstract}

Numerical studies of the gravitational collapse of a stiff ($P=\rho$)
fluid have found the now familiar critical phenomena, namely scaling
of the black hole mass with a critical exponent and continuous
self-similarity at the threshold of black hole formation.  Using the
equivalence of an irrotational stiff fluid to a massless scalar field,
we construct the critical solution as a scalar
field solution by making a self-similarity \ansatz. 
We find evidence that this solution has
exactly one growing perturbation mode; both the mode and the critical
exponent, $\gamma\simeq 0.94$, derived from its eigenvalue agree with
those measured in perfect fluid collapse simulations.  We explain why
this solution is seen as a critical solution in stiff fluid collapse
but not in scalar field collapse, and conversely why the scalar field
critical solution is not seen in stiff fluid collapse, even though the
two systems are locally equivalent.

\end{abstract}
%%%%%%%%%%%%%%%%%%%%%%%%%%%%%%%%%%%%%%%%%%%%%%%%%%%%%%%%%%%%%%%%%%%%%%%%%%%


%%%%%%%%%%%%%%%%%%%%%%%%%%%%%%%%%%%%%%%%%%%%%%%%%%%%%%%%%%%%%%%%%%%%%%%%%%%
\section{Introduction}
%%%%%%%%%%%%%%%%%%%%%%%%%%%%%%%%%%%%%%%%%%%%%%%%%%%%%%%%%%%%%%%%%%%%%%%%%%%

Solutions of Einstein's equations that lie at the threshold of
black-hole formation have provided a unique insight into gravitational
dynamics during the past decade.  The advent of sophisticated
numerical techniques and powerful computers has facilitated the
study of this regime.

Critical phenomena were first discovered in the gravitational collapse
of a real, massless scalar field, in the evolution of generic
1-parameter families of asymptotically flat initial
data~\cite{choptuik_m:1993}. During the time evolution, the matter
either disperses (subcritical data) or collapses to form a black hole
(supercritical data).  The critical solution exists precisely at the
threshold of black hole formation.  This
solution has several interesting
properties which also are at least approximately shared by near-critical 
solutions.  First, the masses of black holes formed in supercritical
evolutions obey a universal scaling law, 
\begin{equation}
\label{equation:mass_scaling}
M_{\mathrm{BH}} \propto |p-\pstar|^\gamma,
\end{equation}
where $p$ is the parameter of a given 1-parameter family of data, and
$\pstar$ is its critical value such that black holes are formed for
$p>\pstar$.  Thus, in this model, infinitesimally small black holes
can be formed by adjusting $p$ to be sufficiently close to $\pstar$.
Second, near-critical evolutions go through a universal intermediate
phase that exhibits self-similar echoing near the origin.  The echoes
are periodic in logarithmic time, so that the critical solution
(universal intermediate attractor) is, in fact, {\em discretely}
self-similar.  Empirically, it is found that the critical exponent,
$\gamma\simeq0.374$, is the same for all 1-parameter families of
massless scalar field initial data.

As they have now been found in a wide variety of general relativistic
self gravitating systems, critical phenomena are considered to be
generic features of ``tuned'' gravitational collapse. Interested
readers are directed to~\cite{critreview2} for a comprehensive review.
In brief, the threshold of black hole formation singles out
well-defined critical solutions which typically have additional
symmetry (beyond that which might have been imposed via the
formulation of the model under consideration), and which, though
unstable by construction, tend to have {\em precisely one unstable
mode} in perturbation theory.  The critical solutions are typically
unique (up to a dynamically irrelevant overall rescaling): the
unstable mode is consequently also unique, as is the growth factor
associated with the mode. The growth factor can then be immediately
related, via ideas familiar from renormalization group studies, to a
scaling exponent defined, for example,
via~(\ref{equation:mass_scaling})~\cite{EvansColeman,koike_t}.

The critical solutions which have been observed so far have either a
time-translational or a scale-translational
symmetry~\cite{lechner_et_al:2002,note_1}; these two types have also
been dubbed Type I and Type II, respectively.  Within each of the two
symmetry categories, solutions can be further differentiated according
to whether the extra symmetry appears continuously or discretely.
Thus, Type I solutions are static or periodic, and display a mass gap
at threshold.  As one tunes closer and closer to criticality, the
static/periodic intermediate attractor (star-like solution) persists
for a lifetime $\tau \sim \sigma \ln \vert p - \pstar \vert$ where
$\sigma$ is the reciprocal Lyapunov exponent associated with the
unstable mode.  Type II solutions, on the other hand, are either
continuously or discretely self-similar (CSS or DSS), exhibit
infinitesimal mass at threshold, and generically have naked
singularities at the origin in the precisely critical limit.  In the
supercritical case, the black hole mass scales according to the
relation~(\ref{equation:mass_scaling}), where again, the scaling
exponent $\gamma$ is the reciprocal Lyapunov exponent of the critical
solution's single unstable mode.  The remainder of this paper concerns
{\em only} Type II behavior, and the key distinction between the two
solutions which will be discussed is that one is CSS while the other
is DSS.  We also note that a key feature of the Type II solutions
which have been computed thus far via direct tuning of PDE solutions
is that, virtually by definition of existing at the threshold of black
hole formation, they do {\em not} contain apparent horizons.

One of the models in which critical phenomena have been studied most
extensively is that of spherically symmetric collapse of perfect
fluids with the one-parameter scale-free equation of state (EOS)
\begin{equation}
P=k\rho.
\end{equation}
Here $P$ is the fluid's isotropic pressure, $\rho$ the energy density,
and $0 < k \le 1$ is a constant (the single parameter of the EOS).
Critical collapse in this context was first considered by Evans and
Coleman for the specific case $k=1/3$ (radiation equation of
state)~\cite{EvansColeman}. These authors not only found a
continuously self-similar (CSS) critical solution via evolution of
families of initial data, but also constructed that same critical
solution more directly by adopting a CSS \ansatz, and then
solving the resulting boundary-value eigenproblem.  This work was
quickly followed by examination of the stability properties of this
and related CSS fluid solutions within the context of perturbation
theory~\cite{koike_t,maison_d:1996,hara_t:1996}.  Working from the CSS
\ansatz, these studies identified single-mode-unstable CSS solutions
for $0\le k \lesssim 0.89$, indicating that the black hole threshold
solutions were CSS (and thus Type~II) for $k$ in this range.  The
absence of solutions for $k\gtrsim 0.89$ was intriguing, and hinted
that the critical solutions for such cases might have different
properties---in particular, there were suspicions that one or both of
the assumptions of 1) regularity at the origin, and 2) continuous
self-similarity might break down at $k \sim 0.89$.  However, CSS
critical solutions with regular origins for $0.89 \lesssim k \le 1$
{\em were} subsequently found, through fluid
evolutions~\cite{neilsen_d:1998,brady_p:1998mg8}, as well as via
adoption of the CSS \ansatz~\cite{neilsen_d:1998}, with, from the
point of view of the PDE evolutions, no obviously special changes in
solution properties at $k \sim 0.89$~\cite{neilsen_d:1998}.

The existence of a CSS critical solution in the limiting case of the
stiff fluid ($k=1$) was particularly surprising to many, as the
spherically symmetric stiff fluid is formally equivalent to a
spherically symmetric real scalar field.  [The technicalities of this
equivalence are detailed below in Section~\ref{ss:fluid-eqs} for the
stiff fluid, and in Appendix~\ref{appendix:equivalence} for an
arbitrary barotropic equation of state $P=P(\rho)$.]  Thus, one might
have naively expected that the critical solution appearing in one
system would automatically arise as the critical solution in the
other.  However, the DSS solution of scalar field collapse and the CSS
solution of stiff fluid collapse are distinctly different solutions.

Two fairly obvious questions then arise: 1) Why is the CSS critical
solution found in stiff fluid collapse not observed in massless scalar
critical field collapse? 2) Why is the DSS scalar field critical
solution not seen in stiff fluid collapse? The answer to the second
question is relatively trivial, and has been well known in the
critical-collapse community for some time---the DSS scalar field
critical solution corresponds to a perfect fluid with negative (as
well as positive) pressures and densities, and hence {\em cannot}
arise in fluid collapse scenarios where energies and densities, are,
by fiat, constrained to be positive. As discussed in more detail
below, the resolution of the second question is somewhat more subtle,
but in essence amounts to the observation that the CSS critical
solution, when interpreted in the scalar field context, contains an
apparent horizon and therefore cannot and does not sit at the black
hole threshold.  This appears closely related to the phenomena
observed in Lechner~{\em et al's} study of general relativistic
non-linear sigma models~\cite{lechner_et_al:2002}, where the absence
of CSS solutions at black hole threshold above a certain strength in
coupling parameter is correlated with the fact that the CSS solutions
at those coupling values have apparent horizons.

%%%%%%%%%%%%%%%%%%%%%%%%%%%%%%%%%%%%%%%%%%%%%%%%%%%%%%%%%%%%%%%%%%%%%%%%%%%
\section{A continuously self-similar (CSS)
scalar field solution with one growing mode}
%%%%%%%%%%%%%%%%%%%%%%%%%%%%%%%%%%%%%%%%%%%%%%%%%%%%%%%%%%%%%%%%%%%%%%%%%%%

In this section we construct the critical solution for spherically
symmetric stiff fluid collapse using the equivalence between the
scalar field and the stiff fluid to work in the scalar field
variables. We first describe this equivalence between the two matter
models, give the field equations in Bondi coordinates, then restrict
them to a continuously self-similar \ansatz. We then study the
spacetime structure and perturbation modes for two solutions, one of
which is the desired critical solution.

%%%%%%%%%%%%%%%%%%%%%%%%%%%%%%%%%%%%%%%%%%%%%%%%%%%%%%%%%%%%%%%%%%%%%%%%%%%
\subsection{The correspondence between the stiff fluid and the scalar field}
\label{ss:fluid-eqs}
%%%%%%%%%%%%%%%%%%%%%%%%%%%%%%%%%%%%%%%%%%%%%%%%%%%%%%%%%%%%%%%%%%%%%%%%%%%

A stiff fluid is a perfect fluid with equation of state, $P=\rho$.
The stress-energy tensor for the stiff fluid can therefore be written
as
\begin{equation}
\label{eq:pfstresstensor}
T^{ab} = \rho (2u^a u^b + g^{ab}) \; .
\end{equation}
Here $\rho$ is the total energy density of the fluid as measured by an
observer moving with a fluid element, and hence having a normalized
4-velocity $u^a$.  If the 4-velocity is irrotational
($\nabla_{[a}u_{b]}=0$), we can write $u^a$ in terms of the gradient
of a scalar field. (In spherical symmetry the fluid is automatically
irrotational.) We introduce a scalar field $\psi$ with future-pointing
timelike gradient such that
\begin{equation}
\nabla_a \psi = \sqrt{2 \rho}\, u_a,
\end{equation}
where $\nabla_a$ is the metric-compatible covariant derivative.  This
defines $\psi$ up to an additive constant, whose value is irrelevant.
The inverse relation between the fluid and the scalar field is then
given by
\begin{eqnarray}
\label{density}
\rho &= &-\half \nabla_c \psi\nabla^c \psi, \\
u_a & = &{\nabla_a \psi \over \sqrt{ - \nabla_c \psi\nabla^c \psi}}.
\end{eqnarray}
Expressing the stress-energy tensor (\ref{eq:pfstresstensor}) in terms
of $\psi$ gives
\begin{equation}	
T^{ab} = \nabla^a \psi \nabla^b \psi - 
	\half (\nabla_c \psi\nabla^c \psi) g^{ab}.
\end{equation}
The conservation of energy-momentum, $\nabla_a T^{ab}=0$, implies that
$\psi$ satisfies the massless, minimally coupled scalar wave equation
\begin{equation}
\nabla_c \nabla^c \psi =0.
\label{eq:waveqn1} 
\end{equation}
This establishes a local, one-to-one relationship between irrotational
(in particular, spherically symmetric) stiff fluid solutions and
massless scalar field solutions with timelike gradient.  The existence
of massless scalar field solutions that also have spacelike and null
gradients is relevant later when we discuss the relation between
interpolating families of stiff fluid solutions and those of scalar
field collapse.  A more general treatment, applicable for equations of
state of the form $P=P(\rho)$, is given in
Appendix~\ref{appendix:equivalence}.

Although the perfect fluid and scalar field models are related by this
mathematical equivalence, differences remain in their physical
interpretation.  Generally, a fluid is a phenomenological model
derived via thermodynamic considerations of a large number of
particles interacting through elastic collisions.  The fluid density,
$\rho$, and four-velocity, $u^a$, are continuum variables defined in
terms of the fundamental discrete variables in a limiting procedure.
The fluid energy density, for example, is defined in terms of the
total energy in an arbitrary volume, in the limit that the volume
element becomes infinitesimally small.  Thus, a fluid is a continuum
model of a discrete system containing a thermodynamically significant
number of particles; the model is expected to fail as the
large-particle-number limit is violated, $\rho\to 0$.  Indeed, the
fluid equations become singular in this limit.

Given the thermodynamic motivation for perfect fluid models, one
conventionally imposes physical constraints on the fluid's energy
density and four-velocity, namely that $\rho > 0$, and that $u^a$ is
timelike.  The fluid equations may admit solutions in violation of
these constraints, in which case the solution may be discarded, or
matched with other solutions to create a physical spacetime.  One
could argue against accepting these physical constraints, but one
would then be adopting a view of a fluid seemingly at odds with the
usual assumption of an underlying discrete system.

The scalar field, on the other hand, is fundamentally a continuum
model, and does not directly represent the behavior of a discrete
system of particles.  Consequently, its solutions are generally not
subject to the constraints discussed in the previous paragraph.  Thus, 
the assumptions made in constructing the
perfect fluid model allow only a restricted class of scalar field
solutions---those with $\nabla^a\psi \nabla_a\psi < 0$---to be
interpreted as {\em physical} fluid solutions.

%%%%%%%%%%%%%%%%%%%%%%%%%%%%%%%%%%%%%%%%%%%%%%%%%%%%%%%%%%%%%%%%%%%%%%%%%%%
\subsection{Coordinate system and field equations}
%%%%%%%%%%%%%%%%%%%%%%%%%%%%%%%%%%%%%%%%%%%%%%%%%%%%%%%%%%%%%%%%%%%%%%%%%%%

The general spherically symmetric line element can be written in Bondi
coordinates as
\begin{equation}
ds^2 = -g \gbar du^2 - 2 g dudr + r^2 (d\theta^2 + \sin^2\theta d\phi^2)
\; ,
\end{equation}
where $g=g(u,r)$ and $\gbar=\gbar(u,r)$.  The coordinate $r$ has
geometrical meaning in terms of the proper area of $r=$~const.
2-spheres.  We demand that the critical solution be regular at the origin;
elementary flatness at $r=0$ then dictates that
$g(u,0)=\gbar(u,0)$. The relation between the fluid density and scalar
field in these coordinates is
\begin{equation}
\rho = \frac{\partial_r \psi}{2 g} ( 2 \partial_u \psi - \gbar
\partial_r \psi) \; .
\end{equation}
A local mass function $m(u,r)$ is defined by
\begin{equation}
1 - \frac{2 m(u,r)}{r} \equiv g^{ab} \nabla_a r \nabla_b r =
\frac{\gbar}{g} \; ,
\end{equation}
where $r$ is understood to be a function on spacetime.

Looking ahead to the application of the CSS \ansatz,
it is convenient to express the Einstein equations using variables
adapted to a scale symmetry:
\begin{equation}
x=-r/u\; , \qquad\qquad\qquad\tau = -\ln(-u)\;.
\end{equation}
This coordinate change is applied directly to the Einstein 
equations (not to the line element), which then become
\begin{eqnarray}
x\gbar' &=& g-\gbar \; , \label{eq:efe1} \\
x(\ln g)' &=& 4\pi \gammahat^2 \; , \label{eq:efe2}\\
(\gbar / g)\dot{\ } + x (\gbar/g)' &=& { 8 \pi  
(\gammahat + \dot{\psi})} \left[ (x - \gbar) \gammahat + x \dot{\psi}
\right] / g \; . \label{eq:efe3}
\end{eqnarray}
Here a prime ($'$) represents differentiation with respect to $x$,
a dot ($\dot{\ }$) means differentiation with respect to $\tau$, 
and we have also
introduced an auxiliary field, $\gammahat(u,r)$, defined by
\begin{equation}
x\, \psi' = \gammahat \; . \label{eq:sfe1}
\end{equation}
Finally, the scalar wave equation of motion~(\ref{eq:waveqn1}) becomes
\begin{equation}
\label{eq:sfe2}
(\psi+\gammahat)\dot{\ } + (x - {\gbar}/{2}) \gammahat' 
= \frac{\gammahat}{2x} (g - 2x)\; .
\end{equation}
Now that the Einstein equations have been written in terms of the 
new variables, the CSS \ansatz\ will reduce them to a system of ODEs.

%%%%%%%%%%%%%%%%%%%%%%%%%%%%%%%%%%%%%%%%%%%%%%%%%%%%%%%%%%%%%%%%%%%%%%%%%%%
\subsection{Continuously self-similar solutions}
\label{s:solutions}
%%%%%%%%%%%%%%%%%%%%%%%%%%%%%%%%%%%%%%%%%%%%%%%%%%%%%%%%%%%%%%%%%%%%%%%%%%%

Self-similar solutions representing scalar field collapse have been
studied by Brady~\cite{brady_p:1995c} and
Christodoulou~\cite{christodoulou_d:1994}.  We present here only those
details that are essential to our discussion; the interested reader
can consult the above-cited references for further details.

The existence of a homothetic symmetry (continuous scale symmetry) in a
spherical spacetime implies that the metric coefficients $g$ and
$\gbar$ depend only on $x = -r/u$, and that the scalar field is of the
form
\begin{equation}
\psi = \hbar(x) + \kappa \tau\; ,
\end{equation}
where $\hbar$ is a function to be determined, and $\kappa$ is a
positive constant.  Restricting equations~(\ref{eq:efe1})--(\ref{eq:sfe2})
to the self-similar \ansatz\ gives
\begin{eqnarray}
x\, \gbar' &=& g-\gbar \; , \label{eq:ssfe1} \\ 
x
(\ln g)' &=& 4\pi \gammahat^2 \; , \label{eq:ssfe2}\\ 
g-\gbar &=&
4 \pi [ 2 \kappa^2 x - (\gbar-2x)(\gammahat^2+2 \kappa\gammahat)]\;
,\label{eq:ssfe3}\\ 
x \hbar' &=& \gammahat \;
, \label{eq:ssfe5}\\
x (2x - {\gbar}) \gammahat' & = & -2\kappa x +{\gammahat} (g - 2x)\; .
\label{eq:ssfe6}
\end{eqnarray}
These equations are singular when $\gbar=2x$. This corresponds to a
similarity horizon in spacetime, that is a null hypersurface of
constant $x$ coordinate. With a slight abuse of terminology, it is
also a sonic surface of the corresponding stiff fluid.  We rescale the
coordinate $u$ by a constant factor such that $x=1$ when
$\gbar=2x$. Note that this means that the usual convention $g=\gbar=
1$ at the origin $x=0$ does not hold.

In order to be regular in the past, say at $u=-1$, the threshold
solution must be analytic at the origin and at the similarity horizon
$x=1$.  This is impossible for general values of
$\kappa$, but we expect to find analytic solutions for isolated values of
$\kappa$ \cite{EvansColeman}.

Christodoulou~\cite{christodoulou_d:1994} has derived a closed form
solution, satisfying these conditions, when $4\pi \kappa^2 = 1/3$.
This is a Friedmann-Robertson-Walker cosmological solution, and is
\emph{not} the similarity solution observed in numerical simulations
of gravitational collapse. We use it here only to test our numerical
methods on an exact solution.  Using a two point shooting method to
search for analytic solutions at other values of $\kappa$, we find a
second solution when $4\pi\kappa^2 \simeq 0.577$. This solution {\em
does} correspond to the stiff fluid critical solution, as discussed
below in Section~\ref{section:comparison}, but first we discuss its
properties as a massless scalar field solution.

\begin{figure}[tb]
\includegraphics[width=8cm,bbllx=100pt,bblly=200pt,bburx=450pt,bbury=650pt]{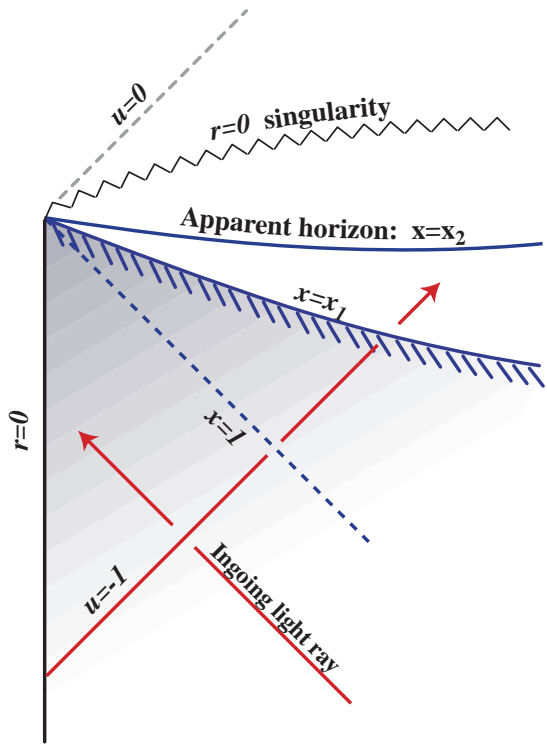}
\caption{\label{fig:st577} A spacetime diagram showing the
self-similar scalar field solution with $4 \pi \kappa^2 \simeq 0.577$.
The apparent horizon, the similarity horizon at $x=1$, and the surface
on which the energy density $\rho$ vanishes are all shown.  The shaded
region, where the scalar field gradient is timelike,
can be identified with a physical stiff fluid solution.}
\end{figure}

The spacetime structure of the $4\pi\kappa^2 \simeq 0.577$ solution is
sketched in Fig.~\ref{fig:st577}.  There is a strong curvature
singularity at $r=0$ and $u=0$; this is a point, in the sense that it
can be surrounded by an arbitrarily small sphere in spacetime.  The
center of spherical symmetry, $x=0$, for $u<0$ is made regular by our
\ansatz, and so is the past light cone of the singularity, $x=1$. The
scalar field gradient is timelike for $0\le x< x_1$, with $x_1 \simeq
1.07$. In that region, the scalar field matter can also be interpreted
as a stiff fluid. The surface $x=x_1$ is a regular spacelike surface
to the future of $x=1$, on which the scalar field gradient is null. In
the fluid interpretation this corresponds to the limit in which the
fluid 4-velocity becomes null, and the comoving fluid energy density
goes to zero.  For $x>x_1$, the scalar field gradient is spacelike,
and the scalar field cannot be interpreted as a stiff fluid given the
usual physical constraints on the scalar field's gradient.  The
spacelike surface $x=x_2 \simeq 1.2$ to the future of $x=x_1$, is an
apparent horizon, i.e., outgoing null rays have zero expansion on this
hypersurface.  Since $x=x_2$ is also a singular point of the
differential equations (\ref{eq:efe1})--(\ref{eq:sfe2}), there is some
concern that the hypersurface may be geometrically singular too.
Nevertheless, since $r$ is finite along $x=x_2$ (except at $u=0$) and
$\gbar/g= 1-2m/r=0$, we conclude that the mass function and all
curvature invariants are finite here, as are the scalar field and its
derivatives (in a regular coordinate system).  Thus, the apparent
horizon is non-singular, and the region to its future is trapped.  We
have not continued the solution to larger values of $x$, but from the
singularity theorems we know that to the future of the apparent
horizon there is a central $r=0$ strong curvature singularity.  This
solution belongs in Class II(a) described in Table~I of
Ref.~\cite{brady_p:1995c}.

%%%%%%%%%%%%%%%%%%%%%%%%%%%%%%%%%%%%%%%%%%%%%%%%%%%%%%%%%%%%%%%%%%%%%%%%%%%
\subsection{Stability analysis}
%%%%%%%%%%%%%%%%%%%%%%%%%%%%%%%%%%%%%%%%%%%%%%%%%%%%%%%%%%%%%%%%%%%%%%%%%%%

Having constructed this CSS scalar field solution, we now turn to an
examination of its stability properties.  As mentioned in the
introduction, we expect a critical solution to have a single unstable
mode, an insight that results from efforts to understand the black
hole mass scaling near the critical
point~\cite{koike_t,maison_d:1996,gundlach_c:1997}.  We therefore
expand each of the fields about its self-similar background value:
\begin{eqnarray}
g &=& g_0(x) + g_1(x,\tau)\; ,\\ \gbar &=&
\gbar_0(x) + \gbar_1(x,\tau)\; ,\\
\gammahat &=& \gammahat_0(x) + \gammahat_1(x,\tau)\; ,\\ \psi
&=& \hbar_0(x) + \kappa \tau+ \hbar_1(x,\tau)\; .
\end{eqnarray}
The subscript $0$ indicates a regular solution of
Eqs.~(\ref{eq:efe1})--(\ref{eq:sfe2}); the
subscript $1$ is used to indicate linear perturbations about this solution.
In terms of these fields, the perturbation equations become:
\begin{eqnarray}
x\, \hbar_1' &=& \gammahat_1 \; ,\label{eq:fofe1}\\
x\, \gbar_1' &=& g_1 - \gbar_1 \; ,\label{eq:fofe2}\\
x\, g_1' &=& 4 \pi \left[g_1\gammahat_0^2+2 g_0\gammahat_0\gammahat_1\right] 
\; ,\label{eq:fofe3}\\
\nonumber
2 x (\gammahat_1+\hbar_1)\dot{\ } &=& x (\gbar_0-2x) \gammahat_1'
+ \gammahat_1 (g_0-2x) \\
&& \mbox{\ \ \ \ }+ 
\gammahat_0 g_1 + \gbar_1 x \gammahat_0' \; .\label{eq:fofe4}
\end{eqnarray}
To satisfy elementary flatness
at the origin, we set $g_1(0,\tau)=\gbar_1(0,\tau)=0$ and
$\gammahat_1(0,\tau)=0$.

We have used two independent numerical methods to search for the
growing modes of Eqs.~(\ref{eq:fofe1})--(\ref{eq:fofe4}), a shooting
method and a matrix eigenvalue method.  Each method has particular
strengths.  Shooting allows accurate determination of the unstable
modes but relies on an initial guess close to the ultimate answer.
For this reason, it is difficult to confirm that all the relevant
modes have been found without further analysis. The matrix eigenvalue
method appears to be less accurate than shooting, but it provides a
scheme to determine all the modes at one time and thus confirm that we have
identified all growing modes.  The observed agreement between the two
methods confirms that there is one unstable mode and two gauge modes.
We describe each method in turn; Table \ref{table:moderesults}
summarizes our numerical results. 

%%%%%%%%%%%%%%%%%%%%%%%%%%%%%%%%%%%%%%%%%%%%%%%%%%%%%%%%%%%%%%%%%%%%%%%%
\begin{table}
\caption{
\label{table:moderesults}
Numerical results for the modes of the continuously self similar
scalar field solution discussed in the text.  Listed in columns 3
through 7 are the mode eigenvalues, $\lambda$, and, where relevant,
corresponding scaling exponents, $\gamma\equiv 1/\lambda$, computed
using the two methods described in text.  Also listed is the scaling
exponent $\gamma_{\rm PDE}$ estimated from  solution of the full
equations of motion~\protect\cite{neilsen_d:1998}.  Quoted uncertainties are
estimates based on the maximum discontinuity at the fitting point for
the shooting method and convergence tests for the matrix method.}

\begin{tabular}{c || d || d | d || d | d || c}

Mode & $\lambda_{\rm exact}$ & $\lambda_{\rm shoot}$ & $\gamma_{\rm shoot}$ & 
       $\lambda_{\rm matrix}$ & $\gamma_{\rm matrix}$ & $\gamma_{\rm PDE}$ \\
\hline
physical unstable & ---   & 1.0654 $\pm$ 0.0005 & 0.9386 $\pm$ 0.0005 & 1.055 $\pm$ 0.01 & 0.95 $\pm$ 0.01 & $\lesssim 0.96$ \\
gauge shift       & 1     & 1.0000 $\pm$ 0.0005 & ---              & 1.013 $\pm$ 0.01 & ---             & --- \\
gauge rescale     & 0     & 0.0000 $\pm$ 0.0005 & ---              & 0.000 $\pm$ 0.00 & ---             & --- \\
\end{tabular}
\end{table}

%physical unstable & --- & $1.065 \pm ?.??$ & $0.939 \pm ?.??$ & $1.055 \pm 0.01$ & $0.95 \pm 0.01$ & $\lesssim 0.96$ \\
%gauge shift & 1.0 & 1.000 & 1.013 \\
%gauge rescale & 0.0 & 0.000 & 0.000

%%%%%%%%%%%%%%%%%%%%%%%%%%%%%%%%%%%%%%%%%%%%%%%%%%%%%%%%%%%%%%%%%%%%%%%%%%%
\subsubsection{Shooting method}
%%%%%%%%%%%%%%%%%%%%%%%%%%%%%%%%%%%%%%%%%%%%%%%%%%%%%%%%%%%%%%%%%%%%%%%%%%%

This method provides direct computation of the eigenmodes of the
perturbation equations (\ref{eq:fofe1})--(\ref{eq:fofe4}) by recasting
them as an eigenvalue problem.  Consider solutions which separate in
the form $f_1(x,\tau)= e^{\lambda \tau}\widehat{f}_1(x;\lambda) $,
where $\lambda$ is a complex number, and the fields $\widehat{f}_1(x;\lambda)$ 
are also complex.  Equations
(\ref{eq:fofe1})--(\ref{eq:fofe4}) then reduce to a set of ordinary
differential equations
\begin{eqnarray}
x\, \widehat{\hbar}_1' &=& \widehat{\gammahat}_1 \; ,\label{eq:sfofe1}\\
x\, \widehat{\gbar}_1' &=& \widehat{g}_1 - \widehat{\gbar}_1 \; ,\label{eq:sfofe2}\\
x\, \widehat{g}_1' &=& 4 \pi \left[\widehat{g}_1\gammahat_0^2
+2 g_0\gammahat_0\widehat{\gammahat}_1\right] 
\; ,\label{eq:sfofe3}\\
\nonumber
2 x \lambda (\widehat{\gammahat}_1 + \widehat{\hbar}_1) &=& 
x (\gbar_0-2x) \widehat{\gammahat}_1' +
\widehat{\gammahat}_1 (g_0-2x) \\
&& \mbox{\ \ \ \ } + 
\gammahat_0 \widehat{g}_1 + \widehat{\gbar}_1 x \gammahat_0' \; .\label{eq:sfofe4}
\end{eqnarray}
Growing modes have real, positive $\lambda$ demonstrating the
existence of an instability.  As explained above,  analyticity of the
solutions at the origin ($x=0$) and the similarity horizon ($x=1$) is
required since the critical solution observed in numerical
simulations must be smooth everywhere.   Analyticity determines the
power series solution about $x=0$ up to three real parameters
corresponding to $\tan^{-1}[(\mathrm{Im}\, \widehat{j}_1 )/
(\mathrm{Re}\, \widehat{j}_1 )]|_{x=0}$, 
$\mathrm{Re} \lambda$ and $\mathrm{Im} 
\lambda$.  The amplitude $|\widehat{j}_1(0)|$ sets the
scale of the perturbations and may be set to unity without loss of
generality.  Similarly,  the power series solution about $x=1$ is determined
up to two complex (four real) parameters--$\widehat{j}_1(1)$ and 
$\widehat{\hbar}_1(1)$.   Equations
(\ref{eq:sfofe1})--(\ref{eq:sfofe4}) were solved using a 4th order
adaptive Runge-Kutta scheme,   shooting from $x=0$ and $x=1$ to a
fitting point at $x=0.5$ until the seven parameters were adjusted to
produce a smooth solution everywhere on the interval.   

There is one physical unstable mode with $\lambda \simeq
1.065$ when $4\pi \kappa^2 = 0.577$.  This suggests that the critical
exponent is $\gamma = 1/\lambda \simeq 0.939$, which is consistent
with the numerical results for stiff fluid collapse where 
$\gamma\lesssim 0.96$~\cite{neilsen_d:1998}. 
In addition to the unstable mode, there 
are modes with $\lambda \simeq 1.000$ and $\lambda \simeq 0.000$.
These modes correspond to the symmetries $u\to u+{\rm const.}$ and
$\psi\to\psi+{\rm const.}$ of the background CSS solutions
respectively, and do not change the physical spacetime. Thus, these
are ``gauge modes'', and finding them at the right place is a (weak)
test of our numerical methods.

For completeness,  we also searched for unstable modes of the CSS
solution with $4\pi\kappa^2=1/3$,  but found only the two gauge modes
as expected.

%%%%%%%%%%%%%%%%%%%%%%%%%%%%%%%%%%%%%%%%%%%%%%%%%%%%%%%%%%%%%%%%%%%%%%%%%%%
\subsubsection{Matrix eigenvalue method}
%%%%%%%%%%%%%%%%%%%%%%%%%%%%%%%%%%%%%%%%%%%%%%%%%%%%%%%%%%%%%%%%%%%%%%%%%%%

As the CSS solution presented above is not the threshold solution in
massless scalar field critical phenomena, it is important to ensure
that we have not missed any other unstable modes. One method to do
this is to write the perturbation equations as $u_{,\tau}=Lu$ and to look
directly for eigenvalues $\lambda$ of the time evolution operator
$L$. 

For this, we have used the background and perturbation equations in
polar-radial coordinates (they are given for example in
\cite{critscalar}), but this is accidental: we could equally well have
implemented the time evolution operator method in Bondi coordinates,
or the shooting method in polar-radial coordinates.

The equations for the linear matter perturbation can formally be
written as the system
\begin{equation}
u_{,\tau}=A(x)u_{,x}+B(x)u+C(x)w, 
\end{equation}
where $u(x,\tau)$ stands for the two first-order matter perturbation
variables and $w(x,\tau)$ stands for the two metric
perturbations. (This is true both for Bondi coordinates and
polar-radial coordinates.)  The metric perturbations are not evolved,
but are reconstructed from the matter perturbations by ODEs of the
form
\begin{equation}
w_{,x}=D(x)w+E(x)u.
\end{equation}
The matrices $A$, $B$, $C$, $D$ and $E$ depend on the background solution.

The perturbation equations were discretized in space but not in
time. The resulting set of equations can formally be written as $d\hat
u_N/d\tau = L_N\hat u_N$. Here $\hat u_N$ is the approximation of $u$
on a grid with $N$ points in the range $0\le x\le 1$, and the matrix
$L_N$ is the corresponding finite difference approximation of the time
evolution operator. The results quoted below were obtained at a
resolution of $N=800$, making $L_N$ a $1600^2$ matrix. The eigenvectors
and eigenvalues of this matrix were then computed to machine precision
using a standard linear algebra package. Most of the eigenvectors of
$L_N$ depend on the finite differencing scheme, but those with the
highest values of ${\rm Re}\lambda$ converge to modes of the continuum
equations with increasing $N$, and the corresponding eigenvalues
converge to the continuum eigenvalues $\lambda$.

The matrix eigenvalue method, in polar-radial coordinates, finds the
growing mode at $\lambda\simeq 1.055$, and the gauge modes at
$\lambda\simeq 1.013$ and $\lambda=0$. This last value is exact to
machine precision as the $\psi\to\psi+{\rm const.}$ gauge mode
decouples in these coordinates.  Using the second-order convergence
of our finite difference scheme, we estimate the uncertainty in
these results to be $0.01$ for our finest grid of $N=800$ points.
This estimate is consistent with the error in $\lambda$ for the gauge
mode, which should be $\lambda=1$, as well as with the independent
results from the shooting method in Bondi coordinates.

With $L_N$ already obtained, it was straightforward to implement the
Lyapunov analysis described by Koike, Hara and
Adachi~\cite{hara_t:1996}, and compare it with our matrix eigenvalue
method. We just had to combine $L_N$ with a fourth-order Runge-Kutta
discretization in $\tau$ in order to obtain a finite difference
version of the evolution equations. (Such a method is variously called
a method of lines, or a semi-discrete method.) We then used the
Lyapunov method to obtain the highest few eigenvalues. In the limit in
which the timestep $\Delta \tau$ is chosen much smaller than the
spatial resolution $\Delta x$, the resulting values of $\lambda$
coincide with those of the direct eigenvector method to machine
precision, as one would expect. However, our method has the crucial
advantage of producing all the top eigen{\it vectors} (modes) as well
as the eigenvalues, whereas the Lyapunov method produces only the one
eigenvector associated with the highest eigenvalue. Finding
subdominant eigenvectors is important here because there are two gauge
modes in the system of equations whose eigenvalues may be higher than
that of the unstable physical mode.

%%%%%%%%%%%%%%%%%%%%%%%%%%%%%%%%%%%%%%%%%%%%%%%%%%%%%%%%%%%%%%%%%%%%%%%%%%%
\subsection{Comparison with the observed stiff fluid critical solution}
\label{section:comparison}
%%%%%%%%%%%%%%%%%%%%%%%%%%%%%%%%%%%%%%%%%%%%%%%%%%%%%%%%%%%%%%%%%%%%%%%%%%%

Having constructed a CSS scalar field solution with a single unstable
mode, we now compare it with the stiff fluid critical
solution.  To facilitate the comparison, we adopt the polar-radial 
coordinates, $(t,r)$, and variables (fluid and metric) 
used in the fluid studies~\cite{hara_t:1996,neilsen_d:1998}.
%{\bf Citations \cite{brady_p:1995c,christodoulou_d:1994} left dangling
%here...}
This is effected by introducing $t = - r \xi(x)$ where
\begin{equation}
x(\ln \xi)' = \frac{\gbar}{\gbar-x} \; , \label{eq:coordxi}
\end{equation}
and $\xi(1)=1$. The metric and fluid variables are
\begin{eqnarray}
a^2 &=& g/\gbar \; ,\\ \omega &=& 4 \pi r^2 a^2 \rho = 2 \pi [ 2 x
(\gammahat^2+\kappa)/\gbar - \gammahat^2]\; ,\\ v^2 &=& \frac{[\gammahat
(\gbar-x) - x \kappa]^2}{x^2 (\gammahat+\kappa)^2} \; .
\end{eqnarray}  

\begin{figure}[tb]
\includegraphics[width=8cm,bbllx=50pt,bblly=50pt,bburx=500pt,bbury=700pt]{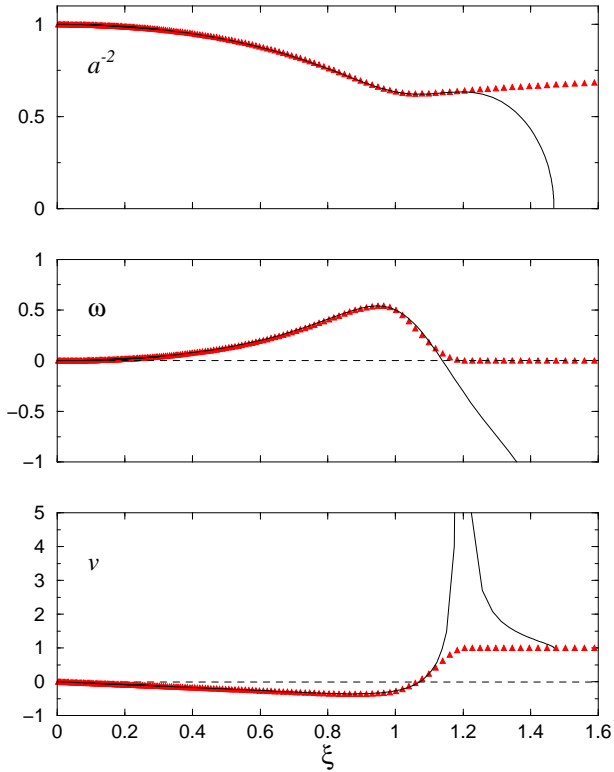}
\caption{\label{fig:577} The CSS solution with $4 \pi \kappa^2 \simeq
0.577$ (solid lines), obtained by solving the ODEs in
Eqs.~(\ref{eq:ssfe1})--(\ref{eq:ssfe6}) and Eq.~(\ref{eq:coordxi}),
compared to a near critical solution computed using the PDE code in
Ref.\protect\cite{neilsen_d:1998} (dotted lines).   The similarity
horizon is at $\xi=1$.  There is excellent agreement between the two
solutions up to, and sightly beyond, the similarity horizon.  The
self-similar scalar field solution cannot be identified with a stiff
fluid solution beyond $\xi\simeq 1.14$ where the gradient of the field
is spacelike.}
\end{figure}

Fig.~\ref{fig:577} shows our CSS scalar field solution and the stiff perfect
fluid critical solution computed by evolving tuned initial data, as
described in Ref.~\cite{neilsen_d:1998a,neilsen_d:1998}. 
There is excellent agreement between the
scalar field solution and the fluid solution inside the similarity horizon
($\xi=1$), and slightly beyond.  The two solutions differ
when $\xi\agt \xi_1\simeq 1.14$, however. 
Here the energy density, $\rho$, of a fluid equivalent to the scalar field
vanishes and the fluid 3-velocity reaches unity.  In the
numerical evolutions of the perfect fluid, a floor is imposed to
maintain $\rho\geq0$ and a timelike 4-velocity in the simulations.
This accounts for the dramatic difference between the solutions at
larger values of $\xi$, and is discussed in detail in Section
\ref{section:CSSinfluid}. 

\begin{figure}[tb]
\includegraphics[width=8cm]{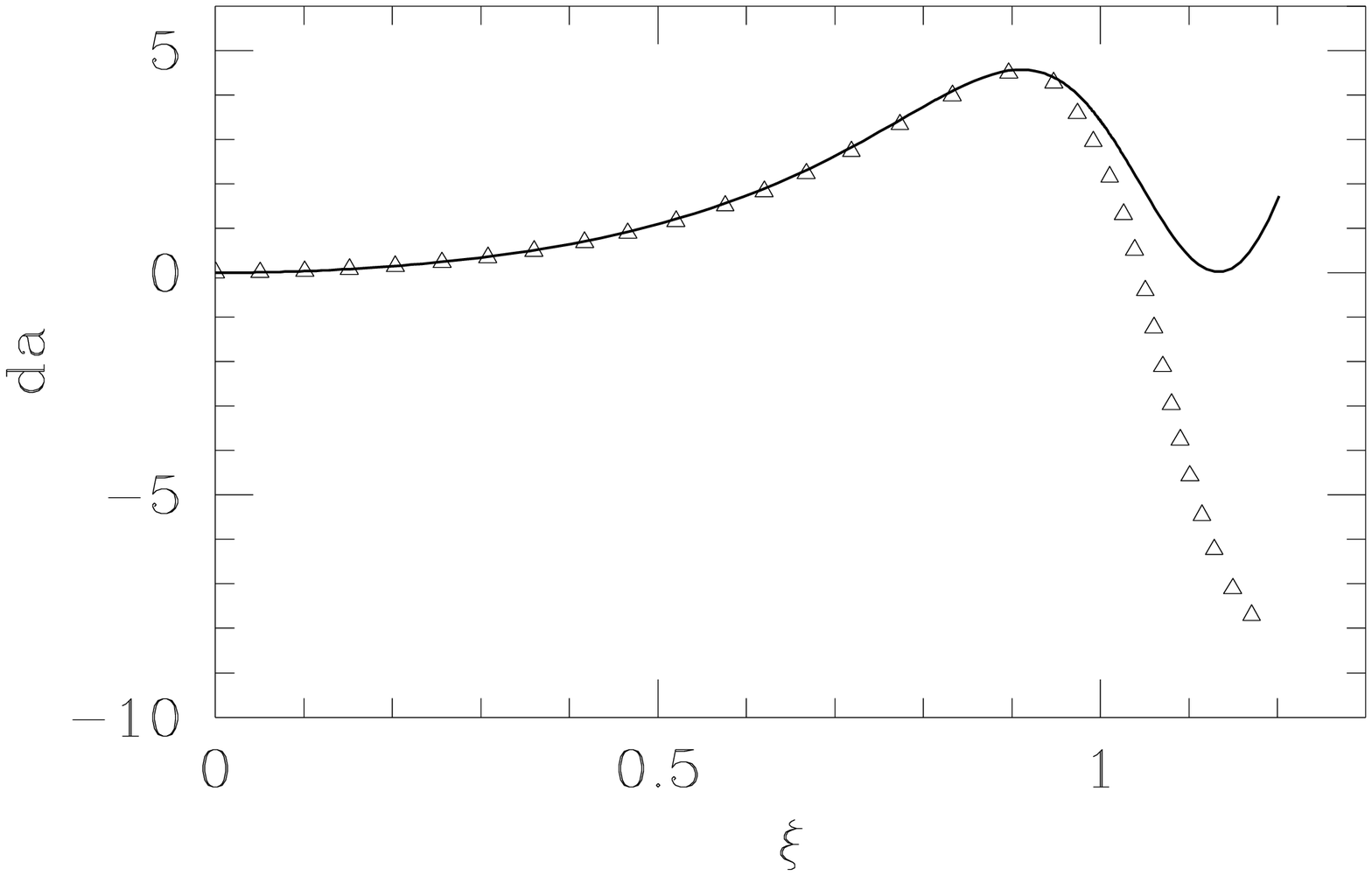}
\caption{\label{figure:eigenmode} This figure shows the 
unstable mode for the metric function $a$, comparing the mode
of the CSS scalar field solution (solid line) with that of the perfect fluid
(triangles).   The modes match particularly well when $\xi \lesssim \xi_1$
($\xi_1 \simeq 1.14$),
where the two solutions are equivalent.  Beyond this point, where the
fluid and scalar field are no longer equivalent, the modes diverge as expected.
The perfect fluid mode was calculated using from 
two near-critical fluid simulations.  As the mode has no intrinsic 
magnitude, the fluid data were scaled to match the scalar field mode.
}
\end{figure}

In addition to this direct comparison of the perfect fluid and scalar
field solutions, we verify that the solutions share the same stability
properties.  As mentioned previously, the unstable mode of the CSS
scalar field solution gives a critical scaling exponent of $\gamma
\simeq 0.94$, which agrees with the perfect fluid estimate of
$\gamma\lesssim 0.96$.  Finally, we can directly compare the unstable
modes as shown in Fig.~\ref{figure:eigenmode}.

%%%%%%%%%%%%%%%%%%%%%%%%%%%%%%%%%%%%%%%%%%%%%%%%%%%%%%%%%%%%%%%%%%%%%%%%%%%
\section{The role of the scalar field CSS solution in gravitational collapse}
%%%%%%%%%%%%%%%%%%%%%%%%%%%%%%%%%%%%%%%%%%%%%%%%%%%%%%%%%%%%%%%%%%%%%%%%%%%

There are at least three regular self-similar, spherically symmetric
scalar field solutions: the cosmological CSS solution found by
Christodoulou, the CSS solution with one growing mode found in stiff
fluid collapse~\cite{neilsen_d:1998} and studied here, and the DSS
critical solution with one growing mode~\cite{choptuik_m:1993}
examined in more detail in Ref.~\cite{gundlach_c:1997}.  Turning to
the questions raised in the Introduction, we examine why the stiff
fluid and scalar field critical solutions are not identical, given the
(local) equivalence between the two matter models.

%%%%%%%%%%%%%%%%%%%%%%%%%%%%%%%%%%%%%%%%%%%%%%%%%%%%%%%%%%%%%%%%%%%%%%%%%%%
\subsection{Self-similar solutions in massless scalar field collapse}
%%%%%%%%%%%%%%%%%%%%%%%%%%%%%%%%%%%%%%%%%%%%%%%%%%%%%%%%%%%%%%%%%%%%%%%%%%%

The cosmological solution ($4\pi\kappa^2=1/3$) has additional
symmetries; it is actually a spatially flat Friedmann solution driven
by an homogeneous scalar field.  It has no growing perturbations and
is therefore a global attractor.  The singularity in this cosmological
solution is velocity dominated. Thus, as one approaches the spacelike
singularity, the evolution of each point in space decouples from that
of neighboring points and evolves approximately as a Friedmann
solution~\cite{AnderssonRendall}.  In general, we expect this solution
to be the final attractor inside the black hole {as the central
singularity is approached}.

The scalar field CSS solution with $4\pi\kappa^2\simeq0.577$ has one
growing mode, and therefore, one might expect it to be a scalar field
critical solution.  However, a critical solution is defined through
its appearance at a black hole threshold, and thus the solution's
global structure must also be considered.  Examining the solution
beyond $x_1$, where the scalar field gradient becomes spacelike, we
find that this solution has an apparent horizon at some $x=x_2$, where
the mass, $m$, and radius, $r$, satisfy $m=r/2$.  This means that the
candidate critical solution is not actually on the black hole
threshold, and that is why it is not a critical solution of scalar
field collapse even though it has precisely one growing mode.

%%%%%%%%%%%%%%%%%%%%%%%%%%%%%%%%%%%%%%%%%%%%%%%%%%%%%%%%%%%%%%%%%%%%%%%%%%%
\subsection{Self-similar solutions in stiff fluid collapse}
\label{section:CSSinfluid}
%%%%%%%%%%%%%%%%%%%%%%%%%%%%%%%%%%%%%%%%%%%%%%%%%%%%%%%%%%%%%%%%%%%%%%%%%%%

\begin{figure}[tb]
\includegraphics[width=8cm]{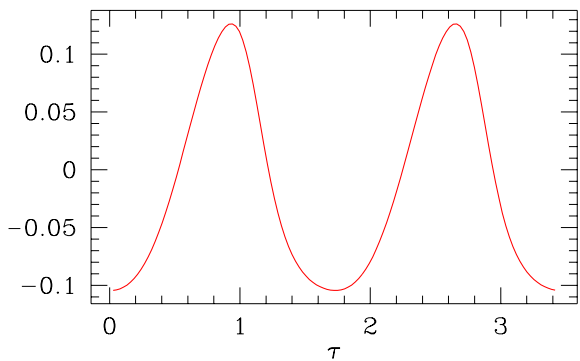}
\caption{\label{figure:density} The equivalent fluid density, given
by Eq. (\ref{density}), of the DSS scalar field solution, plotted for
$0\le\tau\le\Delta\simeq 3.44$ at an arbitrarily chosen point, in this
case $x=0.5$.
}
\end{figure}

We now consider the scalar field CSS solutions discussed above as
candidate solutions for stiff fluid critical collapse.  We can quickly
dispatch with the DSS scalar field solution as a possibility.  As
shown in Fig.~\ref{figure:density}, the equivalent stiff fluid
density, $\rho$, in the DSS solution does not have a definite sign; it
may be positive, zero, or negative for some $\tau$ at any $x$.
Therefore, the scalar field cannot be interpreted globally as a fluid.
As mentioned in the introduction, this has been well known and
understood for some time.

In the previous section, we found that the $4\pi\kappa^2\simeq 0.577$
CSS solution matches the intermediate attractor in stiff fluid
collapse, both in functional form and in the stability properties.
However, this solution is not at the threshold of black hole formation
in scalar field collapse, which raises the immediate question of why
it \emph{is} at the black hole threshold as a stiff fluid.  The answer
to this question comes in demanding that the fluid solution 
(putatively equivalent to the scalar field) be physical.

\begin{figure}
\includegraphics[width=8cm]{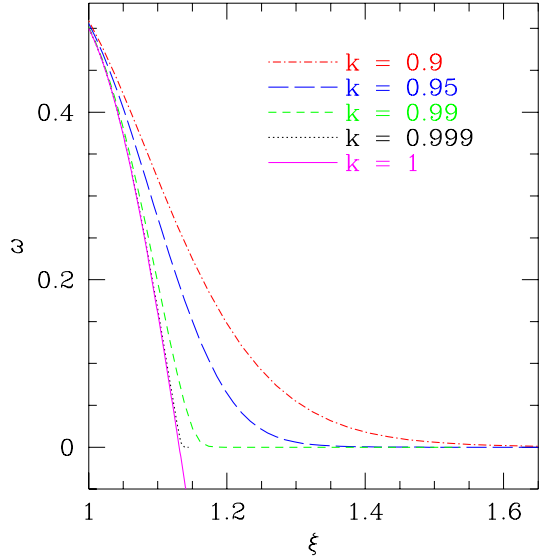}
\caption{\label{figure:omegadetail} A detail of the region about
$\xi_1$ showing $\omega$ for the perfect fluid CSS solutions (obtained
from a CSS \ansatz), including $k \leq 1$.}
\end{figure}

Consider the CSS perfect fluid critical solutions obtained from solving 
a CSS ansatz  for different values of $k$. (See Hara, Koike and
Adachi~\cite{hara_t:1996} for a derivation of the equations, and a
solution method.  See~\cite{neilsen_d:1998} for the solutions with
$k\gtrsim 0.89$.)  Figure~\ref{figure:omegadetail} shows the fluid
density $\omega(\xi)$ as $k\to 1$.  For $k<1$, $\omega$ is smooth,
monotonically decreasing, and asymptotes to zero from above. As $k\to
1$ from below, the curve $\omega(\xi)$ develops a corner joining two
approximately linear sections. In the limit $k=1$ one would expect
$\omega(\xi)$ to reach zero at finite $\xi$ with finite nonzero slope, and
to continue as $\omega=0$ for $\xi>\xi_1$, with the corner becoming a kink
at $\xi_1$. But this is not the case. For $\xi<\xi_1$, the scalar field
solution is indeed the limit $k\to 1$ of the $k<1$ fluid solutions,
but for $\xi>\xi_1$ it is qualitatively different: it continues smoothly
through $\xi_1$ to negative values of $\omega$.

Mathematically, it is a matter of definition how one continues stiff
fluid solutions to the future of points of zero density.  One choice
is to evolve the equivalent scalar field, which remains well-defined
through this transition region.  The density defined in this way
becomes negative but is smooth everywhere. The 4-velocity is not
smooth where the density changes sign, but the three-velocity $v$ with
respect to constant $r$ observers is: it continues smoothly to values
$|v|>1$.  A second choice, motivated by the desire for a ``physical''
fluid solution, is to truncate the CSS solution at $x=x_1$, and
replace the spacetime outside this point with another solution of the
Einstein equations.  It is not possible, however, to match to a vacuum
Schwarzschild solution without invoking some sort of thin shell since
the $T^{00}$ component of the stress-energy tensor does {\it not}
vanish at $x_1$ although the (equivalent) fluid energy density
does. The divergence of the Lorentz factor, $(1-v^2)^{-1/2}$, at $x_1$
keeps $T^{00}$ finite as $\rho\to 0$.

In the numerical simulations~\cite{neilsen_d:1998} of stiff fluid
critical collapse, the fluid solution is also modified for $x\ge x_1$.
During the evolution the code enforces the physical fluid conditions
by imposing a floor, or minimum relative value of the fluid energy
density to the momentum, on the solution.  (See
Appendix~\ref{appendix:pf_sf_vars} for information regarding the
floor, and Ref.~\cite{neilsen_d:1998a} for additional information on
this code.)  The floor is a common, {\it ad hoc} method for allowing
vacuum or near-vacuum regions to develop in numerical solutions.
Examining a near-critical, stiff fluid evolution, we find that the
floor is only active for $x\gtrsim x_1$.  One visible effect of the
floor on this solution is the slight difference in the evolved fluid
solution with the ODE solution near $x\lesssim x_1$, as shown in
Fig.~\ref{fig:577}.  This modification, however, does not change the
scaling properties of the critical solution.  The eigenvalue spectrum
of a self-similar solution depends only on the solution inside the
past light cone of the singularity, i.e., the region $x<x_1$ for the
stiff fluid solution where the floor is not active.

However the CSS solution for the stiff fluid is modified beyond $x_1$,
the effect is the disappearance of the apparent horizon observed in
the scalar field solution. Thus, this solution can appear at the
threshold of black hole formation in stiff fluid spacetimes, and is
indeed the observed critical solution.  This view is further bolstered
by the apparent continuity of the critical solutions in $k$-space for
fluid collapse with equations of state of the form $P = k\rho,\,\,
0.05 \le k \le 1$~\cite{neilsen_d:1998a}.

%%%%%%%%%%%%%%%%%%%%%%%%%%%%%%%%%%%%%%%%%%%%%%%%%%%%%%%%%%%%%%%%%%%%%%%%%%%
\section{Conclusions and Discussion}
%%%%%%%%%%%%%%%%%%%%%%%%%%%%%%%%%%%%%%%%%%%%%%%%%%%%%%%%%%%%%%%%%%%%%%%%%%%

Scalar fields and perfect fluids have historically been important
models in establishing our understanding of critical phenomena in
gravitational collapse.  The dominant features of critical behavior
are now understood for both models in spherical symmetry.  The real
scalar field critical solution is discretely self-similar, Type~II,
and has a mass-scaling exponent $\gamma \simeq 0.37$.  The perfect fluid
critical solutions are also Type~II, continuously self-similar,
and the stiff ($P=\rho$) fluid's mass-scaling exponent measured in
near-critical simulations is $\gamma \lesssim 0.96$.
These models also share a well known formal equivalence, at least
locally, between irrotational stiff fluids and real scalar fields.
Thus, the very different critical behavior observed in these two
possibly equivalent models poses an interesting question:  Why
is a critical solution for one model not the critical solution in
the other?

At least part of the answer comes in how one defines a fluid.
In the usual construction of the fluid model as the continuum
limit for an underlying discrete system, the fluid satisfies
certain physical conditions, e.g., that the co-moving energy density
is positive, and the four-velocity timelike.  These conditions
limit the scalar field solutions that can be interpreted as fluids.
One such disqualified solution is indeed the DSS scalar field
critical solution, for which the equivalent fluid density becomes negative.

To better understand the stiff fluid critical solution, we have
explicitly constructed it as a boundary value problem, using the
scalar field variables to describe the fluid. This confirms that an
exactly CSS, regular, stiff-fluid solution exists.   It has
exactly one growing perturbation mode, whose inverse Lyapunov exponent
$\gamma \simeq 0.94$ 
is equal, to within estimated numerical error, to the critical exponent 
computed from critical collapse of a stiff fluid.
This CSS solution is not seen as a critical
solution in scalar field collapse because it contains an apparent
horizon, and is therefore not at the black-hole threshold.

In spite of the fact that the complete scalar field solution contains
an apparent horizon, and it cannot be interpreted globally as a
fluid, this CSS solution {\it does} appear in critical collapse of the
stiff fluid.  The apparent horizon of the CSS solution is in a region
of the critical solution where the scalar field gradient is spacelike,
so that the equivalent fluid density is negative.  However, the
solution allows a physical fluid interpretation inside the similarity
horizon, and the collapsing fluid apparently is attracted to this part
of the CSS solution.
% (whether constructed with scalar field or perfect
% fluid variables).

Generic stiff-fluid solutions reach points of zero density and
lightlike 4-velocity in a finite time, at which point the fluid
equations break down. One can either continue the solution as a
massless scalar field or one can replace the unphysical fluid with
another solution of the Einstein equations.  In computational
simulations of fluid critical collapse, the solutions are modified
beyond $x=x_1$ via a floor to maintain a positive fluid density.  The
resulting near-critical spacetimes form a continuous family
parameterized by $k$.  The $k=1$ solution approximates the CSS
solution until just before the spacelike surface ($x=x_1$ in our
coordinates) where the density goes to zero. Beyond that surface, they
are qualitatively different---matter at $x \gtrsim x_1$ is ejected at
almost the speed of light, and no apparent horizon forms.  This allows
the modified CSS solution to function as a critical solution at the
black hole threshold.

Finally, we note the recent work of Harada~\cite{Harada}, who argues that 
an {\em additional} unstable mode (the so called ``kink mode''), is 
present in self-similar fluid solutions with $P=k\rho$, for $k \gtrsim 0.89$.
This mode is characterised by a discontinuity in the derivative of 
the fluid density (taken with respect to the similarity variable) at the sonic 
point.  To date, there has been no evidence in near-critical fluid 
evolutions of such additional unstable modes, although if the growth 
factors are sufficiently small, it is entirely possible that they would 
simply not be visible given the numerical precision typically used in 
the PDE evolutions.  In the context of the current work, our requirement 
of analyticity at the sonic point eliminates any possibility of our finding a 
kink mode via the CSS \ansatz.  Further work will be needed to 
clarify this situation.

%%%%%%%%%%%%%%%%%%%%%%%%%%%%%%%%%%%%%%%%%%%%%%%%%%%%%%%%%%%%%%%%%%%%%%%%%%%
\acknowledgments
%%%%%%%%%%%%%%%%%%%%%%%%%%%%%%%%%%%%%%%%%%%%%%%%%%%%%%%%%%%%%%%%%%%%%%%%%%%

We would like to thank J. M. Mart\'\i n-Garc\'\i a for providing help
with implementing the matrix eigenvalue method, and Warren Anderson,
Bob Wald, Richard Matzner, and James Vickers for helpful
conversations.  This work was supported by NSF grants PHY-9407194,
PHY-9970821 and PHY-9614726, by NSERC, and by the Canadian Institute
for Advanced Research.   PRB is also partially supported by an Alfred P. 
Sloan Research Fellowship.

%%%%%%%%%%%%%%%%%%%%%%%%%%%%%%%%%%%%%%%%%%%%%%%%%%%%%%%%%%%%%%%%%%%%%%%%%%%
\appendix
%%%%%%%%%%%%%%%%%%%%%%%%%%%%%%%%%%%%%%%%%%%%%%%%%%%%%%%%%%%%%%%%%%%%%%%%%%%

%%%%%%%%%%%%%%%%%%%%%%%%%%%%%%%%%%%%%%%%%%%%%%%%%%%%%%%%%%%%%%%%%%%%%%%%%%%
\section{Equivalence between an arbitrary fluid and a scalar field}
\label{appendix:equivalence}
%%%%%%%%%%%%%%%%%%%%%%%%%%%%%%%%%%%%%%%%%%%%%%%%%%%%%%%%%%%%%%%%%%%%%%%%%%%

An irrotational perfect fluid is equivalent to a scalar field for any
one-parameter equation of state $P=P(\rho)$
\cite{Moncrief,Landau}. The conservation of the perfect fluid
stress-energy tensor
\begin{equation}
T_{ab} = (P+\rho) u_a u_b + P g_{ab}
\end{equation}
is equivalent to the 3+1 equations
\begin{eqnarray}
&& (P+\rho) u^a\nabla_a u^b + h^{ab}\nabla_a P = 0, \\
\label{conservation}
&& u^a \nabla_a \rho + (P+\rho) \nabla_a u^a = 0.
\end{eqnarray}
Here $h_{ab}=g_{ab}+u_a u_b$ is the projector into the 3-space orthogonal to
the fluid 4-velocity, $u^a$. The first equation is a 3-vector equation, and
is commonly known as the
force, or Euler equation. The second equation is a scalar equation
expressing the conservation of mass/energy. 
We now parameterize $\rho$ and $u^a$ together
through a single vector that is {\em not} a unit vector:
\begin{equation}
w^a \equiv h(\rho)u^a \quad \Rightarrow \quad h(\rho)^2 = - w_a w^a.
\end{equation}
If we now choose the function $h(\rho)$ to be a solution of the
ordinary differential equation
\begin{equation}
\label{hdef}
{dh\over h} = {dP\over P+\rho},
\end{equation}
then we find that the Euler equation takes the simple form
\begin{equation}
\label{force2}
w^a h^{bc} (\nabla_a w_c - \nabla_c w_a) = 0.
\end{equation}
(Up to a factor, (\ref{hdef}) defines $h$ to be the enthalpy
per particle of the fluid.) If the fluid is irrotational
\begin{equation}
\nabla_{[a}u_{b]}=0,
\end{equation}
we also have
$h^{ac}h^{bd}\nabla_{[a}w_{b]}=0$. Together with (\ref{force2}),
we then have $\nabla_{[a}w_{b]}=0$, and therefore, $w_a$ can locally
be written as the gradient of a scalar function,
$w_a=\nabla_a\phi$.  Expressed in terms of 
$\phi$ alone, the remaining field equation (\ref{conservation}) becomes
\begin{equation}
\label{nonlin}
\nabla^a \nabla_a \phi +
\left(1-{d\rho\over dP}\right)h^{-2}\nabla^a \phi \nabla^b \phi \nabla_a
\nabla_b \phi = 0.
\end{equation}
Here $h^2\equiv-\nabla^a\phi\nabla_a\phi$, and $d\rho/dP$ is a given
function of $h$, determined by the equation of state $P=P(\rho)$ and
the solution $h=h(\rho)$ of (\ref{hdef}). Using $u^a\equiv
h^{-1}\nabla^a\phi$ and defining the projector $h^{ab}$ as before, and
with the sound speed squared given by $c_s^2\equiv dP/d\rho$, the
characteristics of the scalar field equation can be emphasized by
writing it as
\begin{equation}
\left(- u^au^b + c_s^2 h^{ab}\right)\nabla_a\nabla_b\phi = 0.
\end{equation}

%%%%%%%%%%%%%%%%%%%%%%%%%%%%%%%%%%%%%%%%%%%%%%%%%%%%%%%%%%%%%%%%%%%%%%%%%%%
\section{Stiff fluid and scalar field variables}
\label{appendix:pf_sf_vars}
%%%%%%%%%%%%%%%%%%%%%%%%%%%%%%%%%%%%%%%%%%%%%%%%%%%%%%%%%%%%%%%%%%%%%%%%%%%

Fluids are usually described analytically using the fundamental 
quantities of density, $\rho$ and four-velocity, $u^a$.  Modern numerical 
methods for solving the fluid equations, however,
are designed to exploit conservation properties, and thus
describe the fluid in terms of its conserved properties, e.g., the fluid's 
energy and momentum.  Modelling a dynamic stiff fluid can 
be challenging, and a new set of conservation variables
\begin{eqnarray}
\label{eq:piphi}
\nonumber
\Pi_{\rm fluid} & = &
(au^r)^2\rho+(\alpha u^t)^2 p + (au^r)(\alpha u^t)(\rho+p), \\
\Phi_{\rm fluid} & = &
(au^r)^2\rho+(\alpha u^t)^2 p - (au^r)(\alpha u^t)(\rho+p).
\end{eqnarray}
was found that considerably improves numerical stability and  
performance~\cite{neilsen_d:1998a}.
These variables are linear combinations of components in the stress-energy
tensor, representing energy and momentum, and
both are positive-definite when the fluid 4-velocity is
timelike, and the density and pressure are positive. 
Nevertheless, numerical error  in a computer evolution 
will sometimes result in values
for $\Phi_{\rm fluid}$ and/or $\Pi_{\rm fluid}$ that are negative,
which translates into a negative fluid density.
In order to preserve the correct physics, a ``floor'' is imposed on these 
variables by forcing them to always be greater than a small, positive 
constant, $\delta$.  This floor is applied at every step in the fluid
update as
\begin{eqnarray}
\Pi_{\rm fluid} &\leftarrow& \max(\delta,\Pi_{\rm fluid}),\\
\Phi_{\rm fluid} &\leftarrow& \max(\delta,\Phi_{\rm fluid}).
\end{eqnarray}
$\delta$ must be small enough that its influence on the dynamics
of the physical fluid is minimized, and its effect can, at least partially,
be judged by repeating the simulations of the same initial data while
varying $\delta$.  In the critical collapse study of 
Ref.~\cite{neilsen_d:1998}, the floor
was set to $\delta=10^{-10}$.

Expressing these new fluid conservation variables in terms of
a set of first-order variables for the scalar field provides an
interesting comparison. A natural choice of such variables
\cite{choptuik_m:1993} is
\begin{equation}
\Phi_{\rm scalar}=\varphi_{,r}, \quad \Pi_{\rm scalar} 
= {a\over \alpha}\varphi_{,t}.
\end{equation}
These are related to the fluid variables for the stiff fluid
($P=\rho$) as
\begin{eqnarray}
\nonumber
\Pi_{\rm fluid} & = & {1\over 2}a^{-2}(\Phi_{\rm
scalar} + \Pi_{\rm scalar} )^2, \\
\Phi_{\rm fluid} & = & {1\over 2}a^{-2}(\Phi_{\rm
scalar} - \Pi_{\rm scalar} )^2.
\end{eqnarray}
We see that the fluid variables are essentially squares of the scalar
field variables. The linear scalar field equation (written in first
order form) is equivalent to the nonlinear fluid equations, and can be
recovered by changing to the ``square-root variables''. The fluid
variables are positive by definition, but one of them goes to zero when
the scalar field variables change sign, as 
\begin{equation}
(\phi_{,\mu}\phi^{,\mu})^2=4\Pi_{\rm fluid}\Phi_{\rm fluid}
\end{equation}
Keeping the fluid variables away from
zero prevents such a sign change and qualitatively changes the
solutions.

%%%%%%%%%%%%%%%%%%%%%%%%%%%%%%%%%%%%%%%%%%%%%%%%%%%%%%%%%%%%%%%%%%%%%%%%%%%

\providecommand{\href}[2]{#2}\begingroup\raggedright\endgroup

%%%%%%%%%%%%%%%%%%%%%%%%%%%%%%%%%%%%%%%%%%%%%%%%%%%%%%%%%%%%%%%%%%%%%%%%%%%

\end{document}